\documentclass[11pt,a4paper]{article}

\usepackage{lmodern} 
\usepackage{microtype} 
\usepackage[english]{babel}

\usepackage{color,graphicx,authblk}

\usepackage{amsmath,dsfont,url,amssymb}
\usepackage{slashed,cancel}
\usepackage[usenames,dvipsnames]{xcolor}
\usepackage[colorlinks=true, linkcolor=black, citecolor=ForestGreen, urlcolor=ForestGreen]{hyperref}
\usepackage{mdframed,overpic}

\usepackage[nosort]{cite}

\newcommand{\ed}[1]{{
#1}}

\DeclareMathOperator{\Tr}{Tr}
\renewcommand{\Im}[0]{\operatorname{Im}}

\begin{document}

\title{A Bound on Thermalization from Diffusive Fluctuations}
\author{ \vspace{10pt} Luca V.~Delacr\'etaz \vspace{10pt}}

\affil[]{\!\!\!\!\em Kadanoff Center for Theoretical Physics, University of Chicago, Chicago, IL 60637, USA\!\!\!\!\!\\
\em James Franck Institute, University of Chicago, Chicago, IL 60637, USA \\
\em Department of Theoretical Physics, Universit\'e de Gen\`eve, 1211 Gen\`eve, Switzerland}

\maketitle

\begin{abstract}

The local equilibration time of quantum many-body systems has been conjectured to satisfy a `Planckian bound', $\tau_{\rm eq}\gtrsim \frac{\hbar}{T}$. 
We provide a sharp and universal definition of this time scale, and show that it is bounded below by the strong coupling scale of diffusive fluctuations, which can be expressed in terms of familiar transport parameters. When applied to quantum field theories at finite temperature, this fluctuation bound implies the Planckian bound. The fluctuation bound moreover applies to any local thermalizing system, and we study its implication for correlated insulators, metals, as well as disordered fixed points, where it can be used to establish a lower bound on the diffusivity in terms of the specific heat $D\gtrsim 1/(c_V^{2/d}\tau_{\rm eq})$. 
Finally, we discuss how the local equilibration time can be directly measured in experiments.

\end{abstract}


\thispagestyle{empty}

\pagebreak

\section{Introduction and Summary}

Interacting many-body systems, whether classical or quantum, thermalize. This process generically involves at least two time-scales: the local equilibration time $\tau_{\rm eq}$, at which the system reaches local thermodynamic equilibrium, and the slower global equilibration or Thouless time $\tau_{\rm Th}$ at which conserved densities relax to global thermodynamic equilibrium. While the Thouless time grows with system size%
	\footnote{For example, in a diffusive system $\tau_{\rm Th} = L^2/D$.},
the local equilibration time is an intrinsic quantity characterizing interactions and thermalization in the system, and is relevant for transport and out-of-equilibrium dynamics more generally. In weakly coupled systems with a quasiparticle description, this time scale can be defined as the average mean free time between large-angle scattering events $\tau_{\rm eq} \equiv \tau_{\rm scat}$. The latter is inversely proportional to the two-to-two cross-section, and is therefore large at weak coupling: e.g., $\tau_{\rm scat} \propto 1/\lambda^2$ if $\lambda\ll 1$ is a dimensionless coupling parametrizing four-body interactions. The local equilibration time can therefore be parametrically large. It has been conjectured, motivated by experiments and theoretical models, that $\tau_{\rm eq}$ cannot instead be parametrically small, namely that there is a quantum limit to how fast many-body systems can thermalize \cite{qptbook,Zaanen2004,Hartnoll:2021ydi}
\begin{equation}\label{eq_Planckian}
\tau_{\rm eq} \gtrsim \frac{\hbar}{T}\, .
\end{equation}
Making this conjecture sharp requires a non-perturbative definition of $\tau_{\rm eq}$, valid away from weak coupling. We propose such a definition in Sec.~\ref{sec_def}: loosely speaking, one should define $\tau_{\rm eq}$ as the time scale of emergence of many-body diffusion (or, more generally, the appropriate hydrodynamic description of the system). By avoiding a reference to the microscopic mechanism for thermalization, which varies from system to system, this definition can apply universally; indeed, some form of hydrodynamics---broadly defined as the emergent dissipative dynamics of conserved densities---is expected to arise in any local interacting system. This definition can furthermore be made sharp using the effective field theory (EFT) of hydrodynamics.

The local equilibration rate $1/\tau_{\rm eq}$ has similarities with the Lyapunov exponent(s) $\lambda_L$, used to characterize chaos in semiclassical systems \cite{larkin1969quasiclassical,ZASLAVSKY1981157,PhysRevB.54.14423,Roberts:2014isa}: both are parametrically small at weak coupling, and expected to be bounded at strong coupling. However, the local equilibration can be defined away from any semiclassical or large $N$ limit; moreover, even at large $N$ it is more experimentally relevant as it affects standard time-ordered correlators, including linear response. While a sharp bound on the Lyapunov exponent was found \cite{Maldacena:2015waa}, a proof of the conjectured bound \eqref{eq_Planckian} has so far remained elusive. Eq.~\eqref{eq_Planckian} is also closely related to conjectured bounds on transport parameters, such as viscosities or diffusivities \cite{Kovtun:2004de,Buchel:2007mf,Hartnoll:2014lpa}. The Planckian bound Eq.~\eqref{eq_Planckian} is in a sense the most primitive: in Boltzmann transport, transport parameters are often obtained from $\tau_{\rm scat}$. Moreover, transport parameters depend on which hydrodynamic theory one considers---e.g., momentum needs to be at least approximately conserved to define viscosities. Instead, the local equilibration time $\tau_{\rm eq}$ always exists (even in superdiffusive and subdiffusive systems).

In this paper, we argue that the local equilibration time is bounded by the strong coupling scale of hydrodynamics. This is a time scale at which hydrodynamic fluctuations become unsuppressed, so that it is not self-consistent for hydrodynamics to emerge. We focus on systems with diffusion, or diffusive attenuation of sound modes, which are the most common and experimentally relevant dissipative universality classes, but our approach can be applied around any other dissipative fixed points, whether superdiffusive like KPZ \cite{PhysRevA.16.732,PhysRevLett.108.180601}, or subdiffusive as arises in constrained systems \cite{ledwith2019angular,Gromov:2020yoc}. In any of these situations, scaling corrections to the dissipative fixed point are small at late times, but large before a certain time scale. In the simplest case of systems with a single conserved density $\dot n + \nabla\cdot j = 0$ with diffusivity $D(n)$, the bound reads
\begin{equation}\label{eq_mybound}
\tau_{\rm eq}
	\geq \alpha_d\frac{(T\chi)^{2/d}}{4\pi D} \left(\frac{D'}{D}\right)^{4/d}\, , 
\end{equation}
where $\chi = dn/d\mu$ is the static charge susceptibility, $D' \equiv dD(n)/dn$, and $\alpha_d$ is a dimensionless coefficient that only depends on the spatial dimension $d$ ($\alpha_1=\frac14$, $\alpha_2 = 2/3$, $\alpha_3=\frac1{2^{2/3}}$). This bound is classical, and has implications for both classical and quantum systems: we expect it to be close to being saturated in any system with a small number of local degrees of freedom. See Ref.~\cite{Michailidis:2023mkd} for numerics in one-dimensional lattices showing emergence of diffusion at this time scale. 

This bound becomes particularly interesting with partial knowledge about the dependence of diffusivity on the background density, $D(n)$. For example, disordered fixed points feature heat diffusion, with a diffusivity that is constrained by scaling symmetry to depend on temperature as $D(T) = D_o T^{1-\frac2{z}}$, with $z\geq 1$ the dynamic critical exponent of the $T=0$ fixed point. The fluctuation bound \eqref{eq_mybound} can then be read as a lower bound on the diffusivity 
\begin{equation}
D \geq \frac{\alpha_d}{4\pi} \left(\frac{2-z}{z}\right)^{4/d} \frac{1}{c_V^{2/d} \tau_{\rm eq}}\, , 
\end{equation}
where $c_V \equiv d\varepsilon/dT$ is the specific heat of the disordered fixed point. Lower bounds on diffusion constants in thermalizing systems have been conjectured in the past \cite{Kovtun:2004de,Hartnoll:2014lpa}, but have, to the best of our knowledge, resisted controlled analytic proofs. A similar bound holds for relativistic quantum field theories (QFTs) and conformal field theories (CFTs); in that case it can furthermore be fruitfully combined with causality constraints \cite{Hartman:2017hhp} to yield 
\begin{equation}
\tau_{\rm eq}
	\gtrsim \frac{1}{s_o^{1/d}} \frac{\hbar}{T}\, , 
\end{equation}
where the dimensionless entropy density $s_o = s/T^d$ is a measure of the degrees of freedom of the QFT. This therefore establishes the Planckian bound for QFTs, up to a dimensionless number that only depends on equilibrium thermodynamics rather than out of equilibrium physics, and therefore has weak coupling dependence and can be obtained using imaginary time methods.%
	\footnote{For example, the 2+1d Ising CFT has $s_o^{1/2} \approx 0.675$ \cite{PhysRevE.56.1642}
, the $O(2)$ CFT has $s_o^{1/2} \approx 0.947$ \cite{PhysRevE.79.041142}, and the quark-gluon plasma at high temperatures $T\gtrsim \Lambda_{\rm QCD}$ has $s_o^{1/3} \simeq \bigl[\frac{\pi^2}{45} \left(4(N_c^2-1) + 7N_cN_f\right)\bigr]^{1/3}\approx 2.75$ \cite{kapusta2007finite}.} 
These and other applications are further discussed in Sec.~\ref{sec_appli}.

\subsection{Relation to previous work}

Part of the original interest and motivation for the Planckian bound was that it may offer a paradigm to understand the linear in temperature resistivity observed fairly universally in the normal phase of high-temperature superconductors \cite{Zaanen2004,Bruin_Planckian,Hartnoll:2014lpa,Legros2019,Hartnoll:2021ydi}. We slightly distance ourselves from the debate about whether or not a time-scale---sometimes called the transport time---can be meaningfully extracted from a dc resistivity, and whether such a time scale should be subject to a Planckian bound (see, e.g., Refs.~\cite{Poniatowski_2021,Hartnoll:2021ydi}). Instead, our work aims to address the broader question of whether there exists a universal bound, in general in out-of-equilibrium quantum statistical mechanics, on a universally defined local equilibration time. We propose to define the local equilibration time $\tau_{\rm eq}$ as the time scale of emergence of diffusive dynamics; with this definition, we expect the Planckian bound \eqref{eq_Planckian} to hold generally, and we in fact prove it in the case of QFTs and CFTs (we furthermore expect this bound to be approximately saturated by generic strongly interacting QFTs). Returning to experiments, this motivates directly measuring $\tau_{\rm eq}$ in correlated materials---this is discussed in Sec.~\ref{ssec_exp}.

Our approach has similarities with previous attempts to obtain bounds on transport and thermalization from hydrodynamic fluctuations \cite{Kovtun:2011np,Chafin:2012eq,Kovtun:2014nsa,Delacretaz:2018cfk,Delacretaz:2021ufg,Hartnoll:2021ydi}. It is closest to Refs.~\cite{Kovtun:2011np,Chafin:2012eq,Kovtun:2014nsa}, which studied local corrections from hyrodynamic fluctuations to transport parameters. The key difference is that we focus on non-analytic (or non-local) corrections from hydrodynamic fluctuations, which prohibit the emergence of hydrodynamics at arbitrarily early times and therefore naturally lead to a lower bound on $\tau_{\rm eq}$. Instead, the analytic corrections considered in \cite{Kovtun:2011np,Chafin:2012eq,Kovtun:2014nsa} are not sign definite in general \cite{Chen-Lin:2018kfl}, and therefore do not lead to bounds. 
Other perspectives on Planckian bounds include \cite{Lucas:2018wsc,Reimann_2019,Pappalardi:2021rwt,Nussinov:2021fgc,doi:10.1073/pnas.2216241120}.

\section{Defining the local equilibration time}\label{sec_def}

Previous attempts at defining the local equilibration time $\tau_{\rm eq}$ have focused on microscopic definitions. These seek to identify both the microscopic {\em mechanism} for thermalization, and the corresponding time scale. Here we will assume thermalization occurs, and instead use a universal consequence of thermalization---diffusion, or hydrodynamics---to define $\tau_{\rm eq}$ from the bottom up.

\subsection{Absence of exponential decay}\label{ssec_noexp}

Before providing a universal definition for the local equilibration time from hydrodynamics, we comment on a commonly used proxy for $\tau_{\rm eq}$: one selects a simple local operator with exponentially decaying two-point function in the thermal state $\langle \mathcal{O}(t)\mathcal{O}\rangle_\beta \sim e^{-t/\tau_{\mathcal{O}}}$. This would allow for the sharp extraction of one or several local thermalization times. However, such operators do not generically exist, as can be illustrated with an archetypical chaotic one-dimensional spin chain, the tilted-field Ising model
%
\begin{equation}
H
	= \sum_i Z_i Z_{i+1} + g X_i + g'Z_i\, , 
\end{equation}
where $X_j,\,Z_j$ are Pauli matrices at site $j$ and with $g,\,g'\sim 1$ chosen such that the model is not accidentally close to an integrable point. This model has no internal global symmetry, but energy is conserved so that energy density diffuses and therefore decays polynomially in the thermal state
\begin{equation}\label{eq_heat_diffusion}
h_i \equiv Z_i Z_{i+1} + g X_i + g'Z_i \, , \qquad\quad
\langle h_i(t)h_i\rangle_\beta \simeq \frac{c_V  T^2}{\sqrt{4\pi D t}} + \cdots \, \quad \hbox{for } t\gg \tau_{\rm eq}\, ,
\end{equation}
where $c_V  = \frac{d}{dT} \langle h_i\rangle_\beta$ is the specific heat, $\langle\cdot \rangle_\beta \equiv \Tr (e^{-\beta H}\cdot)/\Tr e^{-\beta H}$, and $D$ is the heat diffusion constant.
One may expect polynomial decay only occurs for the energy density because it is conserved; however, all local operators in this model also decay polynomially. Consider for example $X_i$: its late time autocorrelation function is in fact proportional to the energy density autocorrelation function at late times
\begin{subequations}
\begin{align}
\langle X_i (t) X_i\rangle_\beta
	\ \simeq\ 
		\frac{\chi_{Xh}^2}{(c_V  T)^2} \langle h_i(t)h_i\rangle_\beta + \cdots 
	\ \sim \ \frac{1}{\sqrt{t}} + \cdots
	 \, \qquad \hbox{for } t\gg \tau_{\rm eq}\, .
\end{align}
\end{subequations}
The coefficient involves a cross susceptibility $\chi_{Xh}\equiv T \frac{d}{dT}  \langle X_i\rangle_\beta$ measuring the overlap of the operator $X_i$ with the diffusing density $h_i$. It is the first term in a more general effective expansion for the operator valid at late times:
\begin{equation}\label{eq_Xconsti}
X_i
	= \frac{\chi_{Xh}}{c_VT} h(x) + \alpha_1 \nabla^2 h(x) + \alpha_2 [h(x)]^2 + \cdots\, .
\end{equation}
Such operator matching equations between a microscopic operator and all possible composite operators of the effective theory in the same representation of the symmetry group are standard in any EFT, and hydrodynamics is no different. In hydrodynamics, operator matching equations are sometimes called {\em constitutive relations}, especially when the microscopic operator is a current density. As long as one includes all terms allowed by symmetry, organized in an expansion in derivatives and fluctuations, one is guaranteed that the correlation functions involving operators on the left or right of this equation match to arbitrary precision at asymptotically late times. While we have focused on $X_i$, one could write a constitutive relation for any other local microscopic operator, e.g.: $Y_i,\,Z_i,\,X_iZ_{i+1}, \,Y_{i-1}Y_iX_{i+1}$, etc. Of course, it is possible to remove by hand the leading term in the constitutive relation, and therefore construct a microscopic operator with autocorrelation function decaying faster than $1/\sqrt{t}$: consider for example
\begin{equation}
\tilde X_i \equiv X_i - \frac{\chi_{Xh}}{cT} h_i\, .
\end{equation}
By construction, its overlap with the energy density vanishes $\chi_{\tilde X h}=0$. However, the general constitutive relation \eqref{eq_Xconsti} makes it clear that hydrodynamics will still control its autocorrelation function, with the leading behavior coming from the overlap with $h^2$:
\begin{equation}
\begin{split}
\langle \tilde X_i (t) \tilde X_i\rangle_\beta
	&\simeq (\alpha_2)^2 \left(\langle h_i(t)h_i\rangle_\beta\right)^2 + \cdots\\
	&\sim \frac{1}{t} + \cdots \, .
\end{split}
\end{equation}
In the thermodynamic limit, any operator supported on a finite number of sites will therefore have polynomial decay at late times. See Refs.~\cite{Delacretaz:2020nit,Glorioso:2020loc} for further examples and tests of operator matching equations of the form \eqref{eq_Xconsti} in thermalizing systems. This conclusion also holds for extensive operators (sums of local operators over all sites); one notable example is the total current \cite{Alder:1970zza,PhysRevB.73.035113,Michailidis:2023mkd}.

\subsection{Equilibration time as the emergence of hydrodynamics}

The example above shows that hydrodynamics permeates correlation functions at late times in thermalizing systems. A more universal way to define $\tau_{\rm eq}$ is to make use of this fact and identify it with the time scale at which the hydrodynamic description becomes accurate. One can make this precise by noting that fluctuating hydrodynamics provides a systematic expansion in derivatives and fluctuations, and thus also predicts arbitrary {\em corrections} to correlation functions such as \eqref{eq_heat_diffusion} at late times. Understanding when these corrections become small will allow us to understand at what times hydrodynamics can emerge. For the autocorrelation function of a diffusing density, one has the following late time expansion%
	\footnote{When $d$ is even, the corrections in the second line have an additional logarithm: $1/t^{d/2}\to \log(t)/t^{d/2}$.}
\begin{equation}\label{eq_autocorrelation_generalform}
\begin{split}
\langle n(x=0,t)n\rangle
	= \frac{\chi T}{(4\pi D t)^{d/2}} \ \times \ 
	\bigg[\ \  &\quad 1 \  + \ \frac{1}{t} a_{0,1} + \frac{1}{t^2} a_{0,2} +\cdots \\
+ \frac{1}{t^{d/2}} &\left( a_{1,0} + \frac{1}{t} a_{1,1} + \frac{1}{t^2} a_{1,2} +\cdots \right)\\
+ \frac{\log t}{t^{d}} &\left( a_{2,0} + \frac{1}{t} a_{2,1} + \frac{1}{t^2} a_{2,2} +\cdots \right) +\cdots \ \bigg] \, ,
\end{split}
\end{equation}
\ed{as we now explain. First, the leading term is the well-known asymptotic behavior in diffusive systems, $\langle n(t)n\rangle \sim 1/t^{d/2}$, which furthermore implies that charge fluctuations scale as $\delta n \sim 1/t^{d/4}$. Nonlinearities in the hydrodynamic equations then lead to fluctuation (loop) corrections to the asymptotic behavior---since one needs at least two nonlinearities to feed back into the two-point function (see Fig.~\ref{fig_EFT}), the leading fluctuation corrections are suppressed by $\delta n^2 \sim 1/t^{d/2}$, corresponding to the $a_{1,0}$ term above. Another source of corrections comes from higher derivative terms in the diffusive equation, such as $(\partial_t - D \nabla^2)n = \alpha\nabla^4 n+ \cdots$. These come with additional powers of $\partial_t$ or $\nabla^2$ compared to the leading terms, and therefore lead to corrections suppressed by integer powers of $1/t$, as in $a_{0,1}$ and $a_{0,2}$ above. Finally, combining both effects, a generic term in this expansion $a_{\ell,n}/t^{n+\frac12d\ell}$ corresponds to the $\ell$-loop correction at $n$th order in derivatives to the two-point function.  }

The coefficients $a_{\ell,n}$ of each of these power-law corrections are transport parameters which, like the more familiar susceptiblities and conductivities, have Kubo formulas relating them to microscopics (see App.~\ref{app_diff}), and can be measured in numerics or experiments; the first few were observed in numerics on $d=1$ lattice systems in \cite{Glorioso:2020loc,Michailidis:2023mkd}. \ed{Some of these are related to hydrodynamic long-time tails, which were observed in a variety of systems, quantum and classical, continuum and on the lattice (see, e.g., \cite{Alder:1970zza,PhysRevB.73.035113}).}
Note that each such power-law correction defines a time scale
\begin{equation}\label{eq_taus}
\tau_{\ell,n}\equiv \left(a_{\ell,n}\right)^{1/(n+\frac12 d\ell)}\, .
\end{equation}
Hydrodynamics clearly receives large corrections at times $t\sim \tau_{\ell,n}$. It is therefore tempting to define the equilibration time as the largest such time scale $\tau_{\rm eq} \equiv \sup \tau_{\ell,n}$. However, this quantity is expected to diverge because Eq.~\eqref{eq_autocorrelation_generalform} is an asymptotic expansion (like any perturbative expansion in field theory), and $\ell$-loop contributions come with a combinatorial factor $\sim \ell!$. This issue can be avoided by removing the factorial growth by hand 
\begin{equation}\label{eq_proposed_def}
\tau_{\rm eq}
	\equiv \sup \frac{\tau_{\ell,n}}{\ell^{2/d}} \, .
\end{equation}
Of course, in practice, only the first few corrections are expected to be experimentally or numerically relevant, so that a reasonable working definition  could also be, e.g., $\tau_{\rm eq}\equiv \max \left\{\tau_{0,1},\,\tau_{1,0},\,\tau_{2,0}\right\}$\ed{, or to identify $\tau_{\rm eq}$ as the time scale where the two-point function is within a certain threshold of its asymptotic form $|\frac{(4\pi D t)^{d/2}}{\chi T}\langle n(t)n\rangle - 1 |<\epsilon$}. The definition \eqref{eq_proposed_def} simply avoids situations where the first few corrections are fine-tuned to zero with as many microscopic tuning parameters.


\subsection{Strong coupling scale vs.~scale of new physics}

In generic systems---i.e.,~systems not close to integrability and without a large number of local degrees of freedom---the time scales appearing in \eqref{eq_taus} are all expected to be parametrically the same, so that any one of them could be used as a proxy for $\tau_{\rm eq}$. However, when studying less generic systems it is important to use a definition such as Eq.~\eqref{eq_proposed_def}, or, e.g., $\tau_{\rm eq}\equiv \max \left\{\tau_{0,1},\,\tau_{1,0},\,\tau_{2,0}\right\}$, that accounts for both loop corrections and higher derivative corrections, as these can come with parametrically different scales. Indeed, the scale of loop corrections corresponds to the strong coupling scale of the theory:
\begin{equation}
\tau_{\ell,0} \sim \tau_{\rm strong \ \! coupling}\, .
\end{equation}
At this time scale, interactions in the EFT give contributions commensurate with tree-level effects; the EFT is strongly coupled. Instead, the higher derivative corrections $a_{0,n}$ in \eqref{eq_autocorrelation_generalform} come from the gaussian part of the EFT. The associated scale is sometimes called the scale of `new physics'
\begin{equation}
\tau_{0,n}\sim \tau_{\rm new \ \! physics}\, .
\end{equation}
At this time scale, the EFT becomes uncontrolled even if it is still weakly interacting, because the infinite tower of irrelevant higher-derivative terms start to give sizeable corrections to observables. A situation where this time scale is parametrically larger than the strong coupling scale $\tau_{\rm new \ \! physics} \gg \tau_{\rm strong \ \! coupling}$ is in systems with approximate symmetries. These lead to almost conserved hydrodynamic densities, with small decay rate $\Gamma$. At times $t\gg 1/\Gamma \equiv \tau_{\rm new \ \! physics}$, this quantity has decayed so is no longer part of the hydrodynamic description, but it leaves an imprint on the EFT by giving large contributions to its Wilsonian coefficients. See Ref.~\cite{Grozdanov:2018fic} for examples of such systems.

Therefore, even though the largest corrections to hydrodynamics at late times typically come from loop corrections,%
	\footnote{Eq.~\eqref{eq_autocorrelation_generalform} shows that this is the case for diffusive systems in $d\leq 2$. For sonic systems this holds in $d\leq 4$.}
the associated time scale $\tau_{\rm strong\ \! coupling}$ can underestimate the time at which hydrodynamics emerges:  $\tau_{\rm strong\ \! coupling} \lesssim \tau_{\rm new \, physics} = \tau_{\rm eq}$.

\section{The bound}\label{sec_bound}

\begin{figure}
{
\large
\begin{align*}
\frac{d}{d\mu}\langle n(t,x)n\rangle\neq 0 \qquad
	 &\hbox{\huge$\Rightarrow$} & \!\!\!\! \!\!\!\! \!\!\!\!
	\begin{gathered}\includegraphics[height=0.08\linewidth,angle=0,trim={0 0 0 -10 },clip]{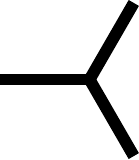}\end{gathered}
    \!\!\!\!
	&&\hbox{\huge$\Rightarrow$} &\!\!\!\!& 
    \!\!\!\!
	\begin{gathered}\includegraphics[height=0.08\linewidth,trim={0 0 0 -10 },clip]{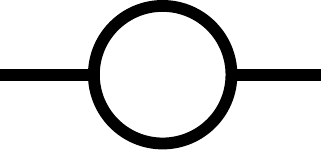}\end{gathered} 
\end{align*}
} 
\caption{\label{fig_EFT} Transport parameters entering in linear response generically depend on densities or their associated potentials. This inevitably introduces nonlinear response $\langle nnn\rangle$ through cubic vertices in the EFT which in turn produce loop corrections to linear response at intermediate times.}
\end{figure}

Many of the coefficients appearing in \eqref{eq_autocorrelation_generalform} correspond to genuinely new transport parameters, i.e.~independent Wilsonian coefficients of EFTs for fluctuating hydrodynamics \cite{Martin:1973zz,Crossley:2015evo}. Therefore, while they can all be measured, they are difficult to constrain analytically in general. This is the case in particular for the higher derivative corrections $a_{0,n}$. Instead, the coefficients of loop corrections $a_{\ell>0,n}$ are typically related to observables at lower loop level such as nonlinear response. In turn, nonlinear response is connected to the dependence of linear response observables to external potentials (see Fig.~\ref{fig_EFT}). The EFT non-trivially ties these various observables \cite{Crossley:2015evo,Chen-Lin:2018kfl,Delacretaz:2023ypv,Michailidis:2023mkd}.

The simplest correction that is fixed in this way is the 1-loop correction to the autocorrelation function \eqref{eq_autocorrelation_generalform} at leading order in gradients, which was shown to be given by \cite{Chen-Lin:2018kfl,Michailidis:2023mkd}
\begin{equation}\label{eq_a10}
a_{1,0}
	=\alpha_d^{d/2}\frac{\chi T}{(4\pi D)^{d/2}} \left(\frac{D'}{D}\right)^2\, ,
\end{equation}
with $D'\equiv dD/dn$ denotes the derivative of the diffusivity with respect to the background value of the diffusing density, and $\alpha_d$ is a numerical coefficient, and $d$ is the spatial dimension. Ref.~\cite{Michailidis:2023mkd} found $\alpha_1= \frac14$;  higher dimensions can be obtained similarly%
	\footnote{Note that the scaling function of the 1-loop correction accidentally vanishes when $d=2$ at $x=0$ -- in this case we evaluate it instead at its maximum $x\simeq \sqrt{Dt}$.}: 
$\alpha_2=2/3$, $\alpha_3=\frac{1}{2^{2/3}}$. As anticipated, the correction $a_{1,0}$ only depends on thermodynamic and transport quantities entering linear response $\chi,\,D$, as well as their dependence on the density through $D'$. In the case of diffusion of energy density $\varepsilon$, the susceptibility is replaced by the specific heat $\chi_{\varepsilon\varepsilon} \equiv T \frac{d\varepsilon}{dT} = Tc_V$ and $D'\equiv dD/d\varepsilon$. This correction is the leading correction to the autocorrelation function in $d=1$ and $d=2$. It was observed numerically in Ref.~\cite{Michailidis:2023mkd}, confirming the EFT prediction including the prefactor.

Given our definition of the equilibration time $\tau_{\rm eq}$ in \eqref{eq_proposed_def}, knowledge of even a single correction provides a lower bound on $\tau_{\rm eq}$. In words, it is not consistent for diffusion to emerge if diffusion itself would predict large corrections at these time scales. The bound reads
\begin{equation}\label{eq_mybound_again}
\tau_{\rm eq}
	\geq \tau_{1,0} = [a_{1,0}]^{2/d} = \alpha_d\frac{(\chi T)^{2/d}}{4\pi D} \left(\frac{D'}{D}\right)^{4/d}\, .
\end{equation}
We explore applications of this bound in Sec.~\ref{sec_appli}.

The coefficient $D'$ can vanish for special values of parameters, in particular in the presence of particle-hole symmetry $\delta n\to -\delta n$. In this case, the leading nonlinearity is a quartic vertex proportional to $D''\equiv d^2 D(n)/dn^2$ and leads to a two-loop correction to the autocorrelation function \cite{Michailidis:2023mkd}
\begin{equation}
a_{2,0} =  \frac{\beta_d^d}{(4\pi D)^{d}}\left(\frac{\chi T D''}{D}\right)^2\, ,
\end{equation}
with $\beta_1=\frac{1}{8\sqrt{2}}$ in $d=1$. The bound then reads
\begin{equation}\label{eq_ddD_bound}
\tau_{\rm eq}
	\geq \tau_{2,0} = [a_{2,0}]^{1/d} = \beta_d\frac{(\chi T)^{2/d}}{4\pi D} \left(\frac{D''}{D}\right)^{2/d} \, .
\end{equation}

The bounds \eqref{eq_mybound_again}, \eqref{eq_ddD_bound} are classical; indeed, the EFT for hydrodynamics is identical for both quantum and classical systems at the first few orders in fluctuations, so that its strong coupling scale has the same dependence on Wilsonian coefficients, or transport parameters. However, these bounds have particularly interesting implications for quantum systems, where transport parameters naturally have factors of $\hbar$. For example, in the finite temperature fan above a quantum critical point described by a CFT, scale and Lorentz invariance imply that diffusivities are given by $D = \frac{\hbar c^2}{T} D_o$, where $D_o$ is a dimensionless number. Similar scalings arise in non-relativistic and disordered fixed points. \ed{Some of these applications require generalizing this bound, and the expansion \eqref{eq_autocorrelation_generalform}, to systems with sound modes in addition to diffusion; this is done in App.~\ref{app_sound} and Sec.~\ref{ssec_CFT}.}

\section{Applications}\label{sec_appli}

\subsection{Disordered fixed points}\label{ssec_disorder}

Disordered fixed points offer one of the simplest application of our bound. Given that momentum is not conserved, their hydrodynamic description at $T>0$ is purely diffusive. Moreover, scale invariance essentially fixes their equation of state and transport parameters: indeed, focusing on fixed points where only energy is conserved for simplicity, the specific heat and thermal diffusivity must take the form
\begin{equation}\label{eq_disordered_scaling}
c_V(T)
	= c_{Vo} T^{d/z}\, , \qquad
D(T)
	= D_o T^{1- \frac2{z}}\, , 
\end{equation}
where $z$ is the dynamic critical exponent of the disordered fixed point%
	\footnote{We note that interacting disordered fixed points are expected to diffuse at $T>0$, regardless of their $T=0$ dynamical critical $z$ (at least as long as $z\leq 2$). We also emphasize that we are identifying $\tau_{\rm eq}$ with the emergence of {\em many-body} diffusion, rather than single-particle quantum diffusion that can arise for free particles in disordered potentials.}.
Here $c_{Vo}$ and $D_o$ are temperature independent constants that only depend on the universality class of the fixed point. See Refs.~\cite{DOROGOVTSEV1980169, PhysRevB.26.154, Thomson:2017dut, Goswami:2017zts, Yerzhakov:2018iwy, Aharony:2018mjm, Goldman:2019xrt, Huang:2023ihu} for controlled constructions of disordered fixed points, working perturbatively in the disorder strength.

The fluctuation bound \eqref{eq_mybound_again} for diffusion of energy density $\varepsilon$, with $\chi_{\varepsilon\varepsilon} = T c_V$, is
\begin{equation}\label{eq_heat_bound}
\tau_{\rm eq}
	\geq \tau_{(1,0)}
	=
	\frac{\alpha_d}{4\pi D} \frac{1}{c_V^{2/d}} \left|\frac{d \log D}{d \log T}\right|^{4/d}\, .
\end{equation}
Using Eq.~\eqref{eq_disordered_scaling}, this becomes
\begin{equation}\label{eq_disordered}
T \tau_{\rm eq} \geq \frac{\alpha_d}{4\pi} \left(\frac{2-z}{z}\right)^{4/d} \frac{1}{D_o (c_{Vo})^{2/d}}\, .
\end{equation}
This already has the flavor of a Planckian bound: the local equilibration time is bounded by the Planckian time, up to however a factor that depends on universal data of the disordered fixed point.

This inequality can also be read as a lower bound on diffusivity:
\begin{equation}\label{eq_D_bound}
D \geq \frac{\alpha_d}{4\pi} \left(\frac{2-z}{z}\right)^{4/d} \frac{1}{c_V^{2/d} \tau_{\rm eq}}\, .
\end{equation}
Universal lower bounds on diffusivities have been conjectured for some time \cite{Kovtun:2004de, Hartnoll:2014lpa, Blake:2016wvh, Trachenko:2020jgr}; to our knowledge this is the first proof of one.%
	\footnote{Instead, upper bounds on diffusivities can be derived from causality or, more generally, speed limits in local systems \cite{Hartman:2017hhp,Lucas:2017ibu,Han:2018hlj}. Certain interesting approaches towards establishing lower bounds include tying transport parameters to microscopics using sum rules (e.g., \cite{Romatschke:2009ng,Nussinov:2021fgc}); these however do not have positive-definite integrands, and have therefore not led to sharp bounds yet. See also Ref.~\cite{Doyon:2019oaf} for bounds on diffusivities in integrable systems.}
Eq.~\eqref{eq_D_bound} applies to any disordered fixed point, we will see in next section that the diffusivity of a CFT can be similarly bounded. It is interesting that the bound \eqref{eq_D_bound} involves a different power of the equilibration time than earlier conjectured bounds \cite{Hartnoll:2014lpa, Blake:2016wvh}: these took the form $D\gtrsim v^2 \tau$, involving a velocity $v$. It is unclear if a disordered fixed point (with $z\neq 1$) has such an intrinsic velocity---instead, the specific heat is a well-defined property of any system where energy is conserved.

For fixed points with $z=1$, where velocities may play a more important role, it is interesting to combine this bound with upper bounds on the diffusivity that follow from causality considerations \cite{Hartman:2017hhp,Roberts:2016wdl,Lucas:2017ibu}
\begin{equation}
D \lesssim v_{\rm B}^2 \tau_{\rm eq}\, , 
\end{equation}
where $v_B$ is the butterfly velocity. Combining this with our bound \eqref{eq_disordered} and dropping numerical factors, this leads to 
\begin{equation}
T \tau_{\rm eq} \gtrsim {\rm max}\left(\frac{1}{D_o c_{Vo}^{2/d}} , \frac{D_o}{v_B^2}\right)
	\geq \frac{1}{v_{\rm B} c_{Vo}^{1/d}}\, .
\end{equation}
The diffusivity has dropped out of the right-hand side of the bound. Establishing a similar Planckian-like bound for $z>1$ disordered fixed points would require `beyond Lieb-Robinson' causality constraints with the appropriate scaling, imposing the vanishing of correlation functions $\langle n(t,x)n\rangle$ for $x\gtrsim t^{1/z}$. Such a constraint would also rule out thermalizing disordered fixed points with $z>2$ (or more precisely: if these thermalize without the help irrelevant corrections, they must be subdiffusive).

We close this section with two additional comments that may motivate future investigation. First, disordered systems have {additional} corrections to correlation functions \eqref{eq_autocorrelation_generalform} due to spatial fluctuations in microscopic transport parameters \cite{Ernst1984}; it would be interesting to incorporate these in the nonlinear EFT of diffusion, and possibly strengthen our bound. Second, our bound Eq.~\eqref{eq_D_bound} takes a simple form if one defines the length scale $\ell_{\rm eq} \equiv \sqrt{D \tau_{\rm eq}}$, so that $c_V^{1/d}\ell_{\rm eq}\geq \left(\frac{\alpha_d}{4\pi}\right)^{1/2} \left(\frac{2-z}{z}\right)^{2/d}$, or, dropping numerical factors, $c_V^{1/d}\ell_{\rm eq}\gtrsim 1$. Similar bounds on an `equilibration length' have been argued for in the past \cite{Delacretaz:2018cfk,Hartnoll:2021ydi}, on the basis that the number of degrees of freedom per thermalizing subsystem (measured here by the specific heat, $c_V \ell_{\rm eq}^d$) must be $\gtrsim 1$. Here we defined $\ell_{\rm eq}$ from equilibration in time \eqref{eq_autocorrelation_generalform}, so it is not clear that it corresponds to the size of the smallest thermalizing subsystem; however it is interesting that our bound is most concisely expressed in terms of this quantity.

\subsection{Quantum field theories}\label{ssec_CFT}

CFTs in spatial dimension $d>1$ placed at finite temperature offer another broad class of thermalizing systems where scale invariance constrains thermodynamic and transport parameters, as in Eq.~\eqref{eq_disordered_scaling} but with $z=1$. These have conserved energy and momentum $\partial_\mu T^{\mu\nu} = 0$, leading to a richer structure of hydrodynamic modes, which can be seen from the retarded Green's function for the momentum density (see, e.g., \cite{Kovtun:2012rj})
\begin{equation}\label{eq_GCFT}
G^R_{T_{0i}T_{0j}}(\omega,q)
	= \left(\delta_{ij} - \frac{q_iq_j}{q^2}\right) \frac{\chi D q^2}{-i\omega + D q^2} - \frac{q_iq_j}{q^2} \frac{\chi c_s^2 q^2}{\omega^2 - c_s^2 q^2 + i \gamma \omega q^2} + \cdots\, ,
\end{equation}
where $\cdots$ denotes higher order terms in $\omega$ or $q$. The first term shows diffusion in the transverse sector. Here $\chi=\varepsilon + P=sT$ is the momentum susceptibility, and the momentum diffusivity is related to the shear viscosity $\eta = \chi D$. The second term shows sound propagation in the longitudinal sector, with velocity $c_s^2 = \frac{dP}{d\varepsilon}$ and sound attenuation rate $\gamma = \frac{2}{d}(d-1)D$. Scale invariance implies $\chi = s_o T^{d+1}$, $c_s^2 = \frac{1}{d}$ and $D = D_o/T$. Similarly, nonlinearities (coming from the Navier-Stokes equations or from the equation of state) are entirely fixed by symmetry---this gives the EFT of hydrodynamics for CFTs a rigid nonlinear structure, reminiscent of nonlinear sigma models, which will lead to certain universal corrections to correlation functions, similar to Fig.~\ref{fig_EFT}.


Fluctuation corrections to hydrodynamic theories with sound modes have been studied since the discovery of long-time tails \cite{Alder:1970zza,PhysRevLett.25.1254,Martin:1973zz, ernst1975nonanalytic, PhysRevA.16.732,andreev1978corrections,morozov1983nonlocal, Kovtun:2003vj,Kovtun:2012rj}. However previous work has either focused on $q=0$ observables, or worked in a mode coupling approximation that does not systematically account for higher order corrections. Understanding thermalization in spacetime requires a full momentum resolved study of fluctuation corrections. This has, to the best of our knowledge, not been done systematically for sound modes before; we do so in Appendix \ref{app_sound}, and use the results below. Near the forward sound front, in the regime $|x| = c_s t + \delta x$, with $\delta x \sim \sqrt{\gamma t}$, we find 
\begin{equation}\label{eq_T00_subleading}
\begin{split}
\langle T_{00} T_{00}\rangle(t,x)
	&= \frac{\chi}{(2\pi \gamma t)^{d/2}} \left(\frac{\gamma}{c_s \delta x}\right)^{d-1}\left[ F_{0,0}^{(d)}\left(\frac{\delta x}{\sqrt{\gamma t}}\right)  + \frac{\tilde a_{1,0}}{t^{(d-2)/2}}   F_{1,0}^{(d)}\left(\frac{\delta x}{\sqrt{\gamma t}}\right) + \cdots \right]\, .
\end{split}
\end{equation}
The first line is the Fourier transform of the leading tree-level correlator \eqref{eq_GCFT}: its first factor is similar to the one in \eqref{eq_autocorrelation_generalform} for diffusion, but the second factor shows an additional suppression $1/\delta x^{(d-1)/2}\sim 1/t^{(d-1)/4}$ due to kinematics of sound, and finally $F_{0,0}^{(d)}$ is a universal scaling function (in $d=3$, it is $F_{0,0}^{(3)}(y) = 2y^2e^{-y^2/2}$). 
\ed{
The second line comes from the leading one-loop correction: its evaluation in Appendix \ref{app_sound} shows that $\tilde a_{1,0} \propto 1/({s D^{(d+2)/2}})$ up to a numerical coefficient that only depends on the dimension $d$.}

Comparing these two terms, one finds an analog of the time scale $\tau_{1,0}$ studied for simple diffusive systems in Sec.~\ref{sec_bound}, which serves as a lower bound on the equilibration time
\begin{equation}\label{eq_CFT_bound}
\tau_{\rm eq} \geq \tilde\tau_{1,0} \equiv \left[\tilde a_{1,0} \right]^{2/(d-2)} 
	=   \frac{1}{D}\frac{\tilde \alpha_d}{(sD^2)^{2/(d-2)}} \, ,
\end{equation}
where $s$ is the entropy density, $D = \eta/(sT)$ the diffusion constant, and $\tilde \alpha_d$ is a numerical factor that only depends on the dimension $d$. This also be read as an upper bound on the diffusivity, or shear viscosity. For example in $d=3$ it reads
\begin{equation}
D T = 
\frac{\eta}{s} \geq \left(\frac{\tilde \alpha_3}{{s_o^2T\tau_{\rm eq}}}\right)^{1/5}\, .
\end{equation}
It is particularly interesting to combine this bound with causality constraints, as in Sec.~\ref{ssec_disorder}. CFTs have a sharp lightcone, leading to an upper bound on the diffusivity $D \lesssim \tau_{\rm eq}$ \cite{Hartman:2017hhp}.%
	\footnote{\ed{This bound is only parametric, i.e.~it is expected to hold up to $O(1)$ coefficients. It is an open question whether there exist sharp causality bounds on transport parameters for generic QFTs.}}
Bounding $\tau_{\rm eq}$ by the maximum of \eqref{eq_CFT_bound} and $D$ produces the Planckian bound 
\begin{equation}\label{eq_Planckian_CFT}
T\tau_{\rm eq}
	\gtrsim \frac{1}{s_o^{1/d}}\, ,
\end{equation}
up to a dimensionless measure of local degrees of freedom of the CFT $s_o = s/T^d$. \ed{The assumption of scale invariance can be lifted without qualitatively affecting the result, provided the bulk viscosity is small $\zeta\ll \eta$ (see Appendix \ref{app_sound}). Eq.~\eqref{eq_Planckian_CFT} can thus be obtained for relativistic QFTs as well, with $s_o$ now a function of temperature. We expect this bound to be parametrically saturated by generic strongly interacting QFTs.} Our arguments apply to $d>1$ QFTs, where hydrodynamic fluctuations are irrelevant, but a similar bound for $d=1$ QFTs was obtained previously \cite{Delacretaz:2021ufg}.

Finally, we note that there is a simplifying limit of QFTs that makes them amenable to stronger bounds: the limit where the local number of degrees of freedom diverges $N\to \infty$ \cite{Adams:2006sv,Arkani-Hamed:2020blm}. In this limit, the hydrodynamic EFT becomes {\em free} (Gaussian). Correlators in the IR then only have simple poles, without the accompanying branch cuts that arise for generic QFTs \cite{Chen-Lin:2018kfl,Delacretaz:2020nit}. This simplified analytic structure leads to strong constraints on thermal correlation functions \cite{Grozdanov:2019kge,Heller:2022ejw,Dodelson:2023vrw} in particular leading to bounds on transport. It is currently unclear whether these approaches are amenable to a systematic $1/N$ expansion. Our approach is in a sense complementary, in that  our bounds become weak at large $N$ (large $s_o$); it would therefore be interesting to combine some of these approaches. For example, Ref.~\cite{Heller:2022ejw} finds bounds in terms of the `radius of convergence' of the gradient expansion of hydrodynamics---while this radius identically vanishes in any theory at finite $N$ due to hydrodynamic fluctuations, our results motivate trying to recast these bounds in terms of the equilibration time $\tau_{\rm eq}$, which can be universally defined. In the context of holographic QFTs, the Planckian bound \eqref{eq_Planckian} is a conjecture about the frequency of the first non-hydrodynamic quasi-normal modes.

\subsection{Measuring $\tau_{\rm eq}$ in experiments}\label{ssec_exp}

Several time scales have been discussed in the context of the Planckian bound, see Ref.~\cite{Hartnoll:2021ydi} for a review. It has nevertheless proven difficult to identify a well-defined time scale that this bound should universally apply to. The transport time $\tau_{\rm tr}$, extracted from the dc resistivity \cite{Bruin_Planckian,Poniatowski_2021}, requires multiplication by an effective mass, and it is not clear which mass should be used in general; it furthermore identifies weakly coupled metals as `Planckian' above the Debye temperature $T_{D}\lesssim T \lesssim E_F$ \cite{Hartnoll:2021ydi,Poniatowski_2021}. The quasiparticle lifetime $\tau_{\rm qp}$ is instead a sharply defined time scale but seems to violate a Planckian bound in certain non-Fermi liquids \cite{Maslov_private}.  The order parameter lifetime \cite{qptbook} assumes exponential decay of bosonic correlators, 
which as discussed in Sec.~\ref{ssec_noexp} does not generically occur. Single particle concepts such as the dephasing time \cite{Altshuler_1982,PhysRevB.65.180202} have also been related to Planckian thermalization. The time scale we propose is perhaps closer to the energy relaxation time $\tau_E$ \cite{Schmid1974}; however, we stress that the definition we have given does not rely on an underlying Boltzmann description.

The universal definition of the local equilibration time $\tau_{\rm eq}$ proposed here, and the possibility that it satisfies a Planckian-like bound, motivates measuring $\tau_{\rm eq}$ in experiments in correlated systems to determine which materials are truly `Planckian'. As discussed in Sec.~\ref{sec_def}, this time scale can easily be measured in numerics, see in particular Ref.~\cite{Michailidis:2023mkd}. \ed{Experiments that access spatial and time resolved near-equilibrium correlators such as Eq.~\eqref{eq_autocorrelation_generalform} can directly extract this time scale as well---we anticipate this to be possible in a number of experimental settings, including: near-field optics \cite{Ni2018}, where sub-micron resolution should furthermore allow to observe time scales relevant for Planckian dissipation; local optical techniques with phase delay, which have been used to measure thermal diffusivity \cite{doi:10.1073/pnas.1703416114}; synthetic quantum systems, where spatial and time resolved probes are particularly accessible \cite{Wienand:2023bkm}. Moreover, momentum and frequency resolved probes such as the dynamic charge response \cite{doi:10.1073/pnas.1721495115} contain the same information as Eq.~\eqref{eq_autocorrelation_generalform}, and can therefore also offer a suitable alternative to direct time-resolved measurements of $\tau_{\rm eq}$.

}

A somewhat more accessible observable in traditional materials is the optical conductivity, which is instead a frequency-resolved probe, at zero wavevector. Similar to the autocorrelation function \eqref{eq_autocorrelation_generalform}, the EFT for diffusion predicts its general form at small frequencies: 
\begin{equation}\label{eq_sigmaw_generalform}
\begin{split}
\sigma(\omega)
	= \sigma_{\rm dc} \ \times \ 
	\bigg[\ \  \quad &1 \  + \ \tilde \tau_{0,1} \omega  \  + \ \tilde \tau_{0,2} \omega +\cdots \\
+ \ \omega^{d/2} &\left( (\tilde \tau_{1,0})^{d/2} + (\tilde \tau_{1,1})^{d/2+1 }\omega +\cdots \right)\\
+ \ \omega^d \log \omega  &\left((\tilde \tau_{2,0})^{d\vphantom{/2}} + (\tilde \tau_{2,1})^{d+1 }\omega +\cdots  \right) +\cdots \ \bigg] \, ,
\end{split}
\end{equation}
with $\omega^{d/2}\to \omega \log \omega$ in $d=2$.%
	\footnote{For all terms in this expansion to arise, it is important that at least two densities, say heat and charge, diffuse; otherwise, the first non-analytic term is $\tilde \tau_{2,1}\omega^{d+1}\log \omega$ \cite{PhysRevB.73.035113}.}
The $\tilde \tau_{\ell,n}$ are analogous to the $\tau_{\ell,n}$ defined in \eqref{eq_autocorrelation_generalform}, so that measuring small corrections to the conductivity at very low frequencies could offer an alternative definition of $\tau_{\rm eq}$. See Refs.~\cite{PhysRevB.56.8714,Marel2003,Delacretaz:2016ivq,PhysRevLett.127.086601,Michon2023} for experiments and theory exploring this possibility in the context of Planckian thermalization. However, the analytic corrections in the first line of \eqref{eq_sigmaw_generalform} are less clearly related to thermalization, as they do not affect late time physics at zero wavevector. \ed{Nevertheless, identifying $\tau_{\rm eq}$ as the inverse frequency where $|\sigma(\omega) - \sigma_{\rm dc}|/\sigma_{\rm dc}$ exceeds a given small threshold would provide a concrete practical definition to use in optics with low ($\lesssim$THz) frequency resolution.} In particular, confirming the observed $\omega/T$ scaling \cite{Delacretaz:2016ivq,Michon2023} at lower frequencies $\omega \lesssim T$ would provide strong evidence that strange metals are indeed Planckian.

\subsection{Heat conduction in crystalline insulators}

In the absence of direct measurements of $\tau_{\rm eq}$, it is interesting to evaluate the right-hand side of the bound \eqref{eq_mybound_again}, which only depends on readily available transport and thermodynamic parameters. This amounts to evaluating the strength of fluctuation corrections to hydrodynamics. We will be particularly interested in situations where these are large, so that the bound becomes most constraining. 

Our results directly apply to heat diffusion in correlated insulators, such as chalcogenides \cite{Littlewood_1980,El-Sharkawy1983,PhysRevLett.101.035901,Pashinkin2009,10.1063/1.4974348,Behnia_2019, PhysRevB.104.035208}. These have also been discussed in the context of Planckian bounds in \cite{doi:10.1073/pnas.1910131116,Behnia_2019,Mousatov:2019vgj}. At high temperatures, i.e. temperatures above the Debye scale $T\gtrsim T_D\sim 200\hbox{ K}$, their diffusivity follows $D \sim v_s^2 \frac{\hbar}{T}$, similar to CFTs but with speed of light replaced by the speed of sound $v_s$. At lower temperatures $T\lesssim T_D$, the specific heat due to acoustic phonon also has a CFT-like form, with $c_V\sim (T/v_s)^d$. Where they meet at $T\sim T_D$, we therefore expect the time scale to be Planckian as in Sec.~\ref{ssec_CFT}. Interestingly, because the specific heat saturates as $T$ increases above the Debye scale (following the Dulong-Petit law), the bound then becomes stronger. Our bound therefore {\em implies} that these systems cannot be Planckian thermalizers at very high temperatures. This is perhaps not entirely surprising, but offers an interesting complementary perspective to previous theoretical investigations. Frequency-resolved experiments showing sub-Planckian time scales in certain insulators support our conclusion 
\cite{PhysRevLett.99.265502,doi:10.1126/sciadv.abg4677}. However, as discussed in the previous section, time-resolved probes would be necessary to unambiguously measure $\tau_{\rm eq}$; see Ref.~\cite{doi:10.1126/science.aav3548} for promising experiments in this direction.

\subsection{Good and bad metals}

The bound Eq.~\eqref{eq_mybound} is loose for Fermi liquids: focusing on heat transport \eqref{eq_heat_bound} with specific heat $c_V \sim {k_F^{d-1}} T/{v_F}$ and thermal diffusivity $D_{\rm th} \simeq \kappa /c_V \sim 1/T^2$ leads to a hydrodynamic strong coupling scale that is parametrically smaller than the expected local equilibration time

\begin{equation}
T\tau_{\rm loop}
	\sim
	\left(\frac{T}{k_F v_F}\right)^{3-\frac2d} \ll
	\frac{k_Fv_F}{T} \sim T\tau_{\rm eq}\, .
\end{equation}
One can check that correlators relevant for charge and thermoelectric transport receive similar corrections (loop corrections for systems with multiple conserved densities were studied in \cite{PhysRevB.73.035113,Michailidis:2023mkd}). Although the bound is loose in this case, we still expect the local equilibration time defined in Sec.~\ref{sec_def} to identify the usual $\sim k_Fv_F/T^2$ time scale of Fermi liquids: this time scale will appear in higher derivative terms $\tau_{(0,n)}$. Thus, while the previously proposed transport time $\tau_{\rm tr}$ incorrectly characterizes good metals as Planckian above the Debye temperature, the definition of $\tau_{\rm eq}$ proposed here identifies them as parametrically sub-Planckian $\tau_{\rm eq} \sim k_F v_F/T^2 \gg 1/T$. For non-Fermi liquids, and bad or strange metals, we similarly expect higher derivative corrections to be the bottleneck for thermalization.

Therefore, unlike for the correlated insulators discussed in the previous section, our bound leaves open the possibility for bad or strange metals to thermalize at a Planckian rate. These have, of course, driven the original interest in Planckian thermalization \cite{Marel2003,Zaanen2004,Bruin_Planckian}; however, as argued in Sec.~\ref{ssec_exp}, time-resolved experiments will be necessary to truly establish whether $\tau_{\rm eq}\sim \hbar/T$ in these systems.

\section*{Acknowledgements}

I am indebted to Kamran Behnia, Erez Berg, Richard Davison, Anatoly Dymarsky, Laura Foini, Paolo Glorioso, Hart Goldman, Sarang Gopalakrishnan, Sean Hartnoll, Cheryne Jonay, Steve Kivelson, Jorge Kurchan, Dmitrii Maslov, Alexios Michailidis,  Silvia Pappalardi, and Misha Stephanov for many discussions, debates and useful feedback that contributed to the ideas presented here. This research was partially supported by the National Science Foundation under Grant No.~PHY-2412710. This work also benefitted from the program ``Quantum Materials With and Without Quasiparticles'' at the Kavli Institute for Theoretical Physics (KITP), which is supported by the National Science Foundation under Grants No.~PHY-1748958 and PHY-2309135.

\appendix

\section{Power law corrections to hydrodynamics from EFTs}\label{app_correction}

Understanding the universal structure of late time thermalization in diffusive systems (Eq.~\eqref{eq_autocorrelation_generalform}), and obtaining expressions for loop corrections used in establishing a bound in Sec.~\ref{sec_bound}, was possible thanks to spacetime resolved expressions for these corrections. These have been obtained only fairly recently for the case of simple diffusion \cite{Chen-Lin:2018kfl,Michailidis:2023mkd} (see also \cite{Jain:2020hcu,Abbasi:2022aao,Lin:2023bli,Chao:2023kvz}), and have been confirmed in numerics \cite{Michailidis:2023mkd} and analytically in models with a perturbative handle is available \cite{Claeys:2021skz}. Analogous calculations have not been performed for systems with sound modes, which include continuum models with momentum conservation, and superfluids.
We will start by briefly reviewing the case of simple diffusion in Sec.~\ref{app_diff}, closely following Refs.~\cite{Akyuz:2023lsm,Michailidis:2023mkd} to which we refer for details, and then generalize this approach to study sound modes in Sec.~\ref{app_sound}.

\subsection{Diffusive systems}\label{app_diff}

An elegant way to understand the emergence of hydrodynamic modes in quantum many-body systems is to view them as Goldstone modes associated with spontaneous breaking of a `doubled' symmetry acting on mixed states \cite{Akyuz:2023lsm,Ogunnaike:2023qyh}, $\rho \to g_1^{-1}\rho g_2$. Indeed, while a pure state may be invariant under such action, a thermal state or canonical density matrix is only invariant if $g_1=g_2$. Focusing first on systems with a single $U(1)$ symmetry, this implies the existence of a Goldstone for the symmetry breaking pattern $U(1)\times U(1)\to U(1)_{\rm diagonal}$. 

To build an EFT for such a system, consider first the situation of spontaneous breaking of all symmetries, $U(1)\times U(1)\to 1$. This describes a system with $U(1)$ symmetry in its ordered phase, e.g.~the XY model for $T<T_c$. The degree of freedom is a superfluid phase for each symmetry, or for each leg of the Schwinger-Keldysh contour, $\phi_{1,2}$. It is convenient to perform a Keldysh rotation and work with the fields
\begin{equation}
\phi_{r} = \frac{1}{2}(\phi_1 + \phi_2)\, , \qquad \quad
\phi_{a} = \phi_1 - \phi_2\, .
\end{equation}
The EFT is then the most general function of $\phi_{r,a}$ invariant under shifts of the Goldstones (see, e.g., Refs.~\cite{altman2013twodimensional,kamenev2023field,Delacretaz:2021qqu,Winer:2022gqz,Donos:2023ibv} for applications of such an EFT). Here we are instead interested in the normal phase, where the symmetry breaking pattern $U(1)\times U(1)\to U(1)_{\rm diagonal}$ only protects the existence of one Goldstone field, $\phi_a$. The system in general contains many other fields, that are invariant (or transform linearly) under the symmetries. In $T=0$ EFTs, such fields are short-lived and can therefore be integrated out without spoiling the locality of the effective action. However, in the Schwinger-Keldysh context, one linear combination of matter fields mixes with the Goldstone in the action at the Gaussian: the density $n_r = \frac12 (n_1 + n_2) \equiv n$. The minimal EFT therefore contains $\phi_a$ and $n$, and is given by
\begin{equation}\label{eq_S_eff}
S_{\rm eff}[n,\phi_a] = 
	\int dt d^d x \, n \dot \phi_a - D(n) \nabla \phi_a\cdot \nabla n + i T \sigma(n) (\nabla \phi_a)^2 + \cdots\, ,
\end{equation}
where $\cdots$ denotes terms with more derivatives or $\phi_a$ fields. The Wilsonian coefficients $\sigma, D$ can depend on the density $n$ in general. Expanding around the desired background value of density $n=\bar n +  \delta n$, the Gaussian action leads to the following propagators at tree-level
\begin{subequations}\label{eq_propa}
\begin{align}
\langle n\phi_a\rangle_0 (\omega,q)
	= \frac{1}{\omega + i D q^2}\, , \qquad\quad
\langle nn\rangle_0(\omega,q)
	= \frac{2T\sigma q^2}{\omega^2 + D^2 q^4} \, ,
\end{align}
\end{subequations}
whose Fourier transform produces the first term in Eq.~\eqref{eq_autocorrelation_generalform}. These correlators will receive corrections from the infinite tower of irrelevant terms in \eqref{eq_S_eff}. The leading loop correction comes from the cubic interaction that arises from expanding $D(n)$ in \eqref{eq_S_eff}: $D(n) = D + D'\delta n + \cdots$ and is given by \cite{Michailidis:2023mkd} (for $d$ odd):%
	\footnote{Note that the loop integral also has a UV divergent piece, that is analytic in $\omega,q$. These UV divergences can be absorbed by counterterms $\delta D$ or $\delta \sigma$ in the EFT. The focus of this paper are IR singular pieces, which are not affected by UV divergences.  }
\begin{equation}\label{eq_1loopGR}
G^R_{nn}(\omega,q)
	= \frac{\chi D q^2}{-i\omega + D q^2} - \frac{\sigma q^4}{(-i\omega+ Dq^2)^2} \frac{\chi D'^2}{D^2}(-i\omega) \frac{\left(\frac{2i\omega}{D} - q^2\right)^{\frac{d}2-1}}{(16\pi)^{d/2}\Gamma(\frac{d}2)} + \cdots\, .
\end{equation}
The Fourier transform of this correction corresponds to the $a_{1,0}$ term in \eqref{eq_autocorrelation_generalform}. 

As for more familiar transport parameters, one can establish a Kubo formula that extracts the coefficient $a_{1,0}$ of this correction. Focusing on $d=1$ for simplicity, and using the Ward identity to obtain the current correlator $G^R_{jj}(\omega,q) = \frac{\omega^2}{q^2}G^R_{nn}(\omega,q)$, one has
\begin{subequations}
\begin{align}
\lim_{\omega\to 0} \lim_{q\to 0} \frac{1}{\omega}\Im G^R_{jj}(\omega,q) 
	&= \chi D = \sigma\, ,\\
\lim_{\omega\to 0} \lim_{q\to 0} \frac{1}{\sqrt{\omega}}\partial_q^2\Im G^R_{jj}(\omega,q)
	&= -\sigma D \sqrt{\pi} a_{1,0}\, .
\end{align}
\end{subequations}
The first line is the usual Kubo formula for conductivity, whereas the second line extracts the 1-loop correction. Alternatively, $a_{1,0}$ can be extracted in time domain from Eq.~\eqref{eq_autocorrelation_generalform} by considering $\lim_{t\to \infty}\partial_t[ t^{d/2}\langle n(t)n\rangle]$. Unlike usual transport parameters, one finds from Eq.~\eqref{eq_1loopGR} that the EFT predicts this correction is {\em fixed} in terms of other transport and thermodynamic parameters: $a_{1,0} = \frac{1}{2} \frac{\chi T}{(4\pi D)^{1/2}} \left(\frac{D'}{D}\right)^2$.

\subsection{Sonic systems}\label{app_sound}

We now turn to systems with momentum conservation in $d>1$ spatial dimensions. Our goal is to obtain the leading intermediate time corrections to hydrodynamic correlators such as \eqref{eq_T00_subleading}. This question has been studied long ago in a mode coupling approximation \cite{ernst1975nonanalytic}, which correctly captures the qualitative features of the leading correction. However, a systematic late time expansion requires the controlled framework of EFT which accounts for all irrelevant corrections to hydrodynamics, including nonlinearities in the `noise field' ($\phi_a$ above). We therefore generalize the EFT for systems with momentum conservation, focusing for concreteness on relativistic QFTs. A similar construction appeared in Ref.~\cite{Jain:2020hcu}, which however considered incompressible fluids and focused on the static limit $\omega\to 0$.

In analogy with the diffusive example from the previous section, the thermal state can be viewed as spontaneously breaking the doubled spacetime translation symmetry down to the diagonal subgroup $\mathbb R^{d+1}\times \mathbb R^{d+1} \to \mathbb R^{d+1}_{\rm diag}$. This leads to a Goldstone for each spacetime translation, that we will denote by $X_\mu^a$ (the analog of $\phi_a$ above). As before, the other dynamical fields in the EFT are the conjugate densities $T^{0\mu}$.%
	\footnote{To make the analogy with the simple diffusive EFT \eqref{eq_S_eff} as clear as possible, Lorentz invariance is not made manifest here (unlike in Ref.~\cite{Jain:2020hcu}). This also allows to straightforwardly generalize Eq.~\eqref{eq_S_eff_fluid} to systems without boost symmetry. Making Lorentz invariance manifest would nevertheless be useful to efficiently push our calculation to higher order corrections. }
Keeping only terms up to cubic order in fields, one arrives at the following EFT:
\begin{align}\label{eq_S_eff_fluid}
S_{\rm eff}[T^{0\mu}, X_\mu^a]\notag
	&= \int dt d^dx \, \Bigl[T^{0\mu}\dot X_\mu^a + T^{0i}\partial_i X_0^a + c_s^2 T^{00}\partial_i X^a_i\\ \notag
	&- \eta \left(\partial_i X_j^a + \partial_j X_i^a\right) \left(\tfrac{1}{\chi}\partial_i T^{0j} - i T \partial_i X_j^a\right) - (\zeta-\tfrac{2}d\eta) \partial_i X^a_i \left(\tfrac{1}{\chi}\partial_j T^{0j} - i T \partial_j X^a_j\right)\\
	&+ \frac{1}{\chi} T^{0i}T^{0j}\partial_i X_j^a + \cdots \Bigr]
\end{align}
The first line contains the Gaussian non-dissipative terms. The coefficient of the first term is 1 by definition of the densities $T^{0\mu}$; for the second term, this is a consequence of Lorentz invariance ($T^{i0}=T^{0i}$). The coefficient $c_s^2$ of the third term is arbitrary, and can be a function of $T^{00}$ and $(T^{0i})^2$. The second line contains the leading dissipative terms, involving the shear and bulk viscosity $\eta$ and $\zeta$. KMS symmetry fixes the $(X^a)^2$ terms. Finally, the third line contains the leading nonlinearity, whose coefficient $\chi = \varepsilon + P = s T$ is also fixed by boost symmetry (as in the Navier-Stokes equation). The Navier-Stokes equations with noise follow from the equation of motion $\delta S / \delta X^a_\mu = 0$ and identifying the velocity $v^i = T^{0i}/\chi$.

We start by considering CFTs, where $c_s^2 = 1/d$ and $\zeta=0$. The only cubic vertex at leading order in derivatives is the one shown in the last line of \eqref{eq_S_eff_fluid}. Using again diffusive scaling $T^{0\mu}\sim X^a_\mu \sim q^{d/2}$, one finds that this cubic term scales as $q^{(d-2)/2}$; in $d>2$, it is therefore irrelevant, allowing for a controlled expansion at low frequencies and momenta.%
	\footnote{In $d=2$, it is only marginally irrelevant. Hydrodynamic EFTs then still offer a controlled framework to capture low frequency and momentum response, although this requires resumming large logarithms \cite{PhysRevA.16.732}.}
The tree-level correlators can be obtained from the Gaussian part of the action, for example
\begin{equation}\label{eq_TX_para_perp}
\begin{split}
\langle T^{0i} X^j_a\rangle_0(\omega,q)
	&= \frac{q^i q^j}{q^2}  \frac{-i\omega}{\omega^2 - c_s^2 q^2 + i\gamma \omega q^2}
	+ \left(\delta_{ij} - \frac{q^i q^j}{q^2}\right) \frac{1}{\omega+iDq^2}\\
	&\equiv \frac{q^i q^j}{q^2} \langle TX\rangle_\parallel 	+ \left(\delta_{ij} - \frac{q^i q^j}{q^2}\right)\langle TX\rangle_\bot\, .
\end{split}
\end{equation}
with diffusivity $D = \eta/\chi$ and sound attenuation rate $\gamma = \frac2d(d-1)D$. In the second line we separated this correlator into the longitudinal channel $\langle TX\rangle_\parallel$, which carries the sound mode $\omega\simeq \pm c_s q - \frac{i}2 \gamma q^2$, and the transverse momentum channel $\langle TX\rangle_\bot$ which is diffusive $\omega = -i Dq^2$. The 1-loop correction to the stress tensor correlator is given by
\begin{equation}\label{eq_TT_loop}
\delta \langle T^{0k} X^l_a\rangle(p)
	= -\frac{1}{\chi^2}
	\langle T^{0k} \partial_{(i}X_{j)}\rangle(p) \langle T^{0j'} X^l\rangle(p)
	\int_{p'} \langle T^{0i} T^{0i}\rangle(p-p') \langle T^{0j}\partial_{(i'}X_{j')} \rangle(p')\, , 
\end{equation}
where $p$ denotes frequency and wavevector $p=\{\omega,q\}$ and $\int_p \equiv \int \frac{d\omega d^dq}{(2\pi)^{d+1}}$. We omit the $a$ subscript on $X_a^i$ for clarity.

We will focus on fluctuation corrections to sound, $\langle TX\rangle_\parallel$. One of the leading contributions in \eqref{eq_TT_loop} arises from diffusive propagators in the loop, and takes the form
\begin{equation}
\delta \langle T X\rangle_\parallel(p)
	= - [\langle T X\rangle_\parallel(p)]^2 \frac{4}{\chi^2 q^2}\int_{p'} \left(q\cdot \Delta_{q-q'}\cdot q'\right)\left(q\cdot \Delta_{q'}\cdot q\right) \langle TX\rangle_\perp(p') \langle TT\rangle_\perp(p-p')\, , 
\end{equation}
and we defined the projector $\left(\Delta_q\right)_{ij}\equiv \delta_{ij} - \frac{q_iq_j}{q^2}$. One of the legs in the loop involves the transverse stress tensor two-point function  $\langle TT\rangle_\perp(p) = \frac{2T \eta q^2}{\omega^2 + D^2 q^4}$, whereas $\langle TX\rangle_\perp$ was defined in \eqref{eq_TX_para_perp}. The integral can be evaluated for general kinematics, but we are most interested in the late time corrections \eqref{eq_T00_subleading}, which come from corrections near the sound pole $\omega\simeq c_s q$. In this kinematics, the integral over $q'$ is dominated by fairly large momentum $q' \sim \sqrt{c_s q/D}$. Taking $q'\gg q$ in the integrand, one finds
\begin{equation}
\delta \langle T X\rangle_\parallel(p)
	= - [\langle T X\rangle_\parallel(p)]^2 \frac{4T}{\chi q^2} \int \frac{d^{d-1}\hat n}{(2\pi)^d} (q^2 - (q\cdot \hat n)^2)^2 \int_0^\infty \frac{dq' \, q'^{d-1}}{\omega + 2iDq'^2}\, ,
\end{equation}
where the first integral is over the unit sphere, parametrized by $\hat n$. Specializing to $d=3$, and absorbing the analytic UV divergence with a counterterm, one is left with a singular piece
\begin{equation}\label{eq_sdds}
\delta \langle T X\rangle_\parallel(p)
	= [\langle T X\rangle_\parallel(p)]^2 \frac{\omega^{1/2}}{(2iD)^{3/2}} q^2  \frac{8}{15\pi} \frac{T}{\chi } \, .
\end{equation}
Evaluated near the sound pole $\omega \simeq c_sq$, this self-energy $\Sigma\sim q^{5/2}$ leads to a correction to the sound dispersion relation $\omega = c_s q - \frac{i}2 \gamma q^2 + \#\frac{T}{\chi D^{3/2}}q^{5/2}$, which upon Fourier transformation produces the corrections anticipated in Eq.~\eqref{eq_T00_subleading}.

We comment on the generalization to QFTs without scale invariance. The diffusive loop contribution to sound Eq.~\eqref{eq_sdds} is unchanged. The part of the scaling function $F_{1,0}^{(d)}$ in \eqref{eq_T00_subleading} arising from diffusive loops is therefore the same, except for the overall coefficient $\tilde a_{1,0}$ which has an extra factor of $(\gamma/[\frac2d(d-1)D])^{(d-2)/2} = (1+\frac{d}{2(d-1)}\frac{\zeta}{\eta})^{(d-2)/2}$. This factor is close to one if $\zeta\ll \eta$, in particular when the QFT is close to a CFT. Finally, the non-trivial equation of state of QFTs produces a second cubic vertex in \eqref{eq_S_eff_fluid}, because $dc_s^2 / d\varepsilon \neq 0$. This leads to an additional sound loop contribution to $\delta \langle T X\rangle_{\parallel}$; we leave its evaluation for future work.


\bibliographystyle{ourbst}
\bibliography{ref_flucbound}{}

\end{document}